\documentclass[conference]{IEEEtran}

\IEEEoverridecommandlockouts\usepackage{cite}
\usepackage{amsmath,amssymb,amsfonts,mathtools} 
\usepackage{algorithm, algpseudocode}
\usepackage{graphicx}
\usepackage{textcomp}
\usepackage{xcolor}
\usepackage{bm}
\usepackage{float}
\usepackage{soul}
\usepackage[acronym,shortcuts]{glossaries}
\usepackage{outlines}
\usepackage{algorithm}
\usepackage[english]{babel}
\usepackage{blindtext}
\usepackage{subcaption}
\usepackage{orcidlink}
\usepackage{lipsum}
\usepackage{stfloats}
\usepackage[]{hyperref}

\def\BibTeX{{\rm B\kern-.05em{\sc i\kern-.025em b}\kern-.08em
T\kern-.1667em\lower.7ex\hbox{E}\kern-.125emX}}

\newacronym{PDF}{PDF}{probability density function}
\newacronym{CR}{CR}{cognitive radio}
\newacronym{SotA}{SotA}{state of the art}
\newacronym{OFDM}{OFDM}{orthogonal frequency division multiplexing}
\newacronym{PAPR}{PAPR}{peak-to-average power ratio}
\newacronym{IFFT}{IFFT}{inverse fast Fourier transform}
\newacronym{DFT}{DFT}{discrete Fourier transform}
\newacronym{SQNR}{SQNR}{signal to quantization noise ratio}
\newacronym{ADC}{ADC}{analog to digital converter}
\newacronym{DAC}{DAC}{digital to analog converter}
\newacronym{TR}{TR}{tone reservation}
\newacronym{RT}{RT}{reserved tone}
\newacronym{PRT}{PRT}{peak-reserved tone}
\newacronym{LP}{LP}{linear programming}
\newacronym{SLM}{SLM}{selective mapping}
\newacronym{DVB-T2}{DVB-T2}{DVB-second generation}
\newacronym{DVB-NGH}{DVB-NHG}{DVB for next generation Handled}
\newacronym{QCQP}{QCQP}{quadratic constrained quadratic program}
\newacronym{FP}{FP}{fractional programming}
\newacronym{CCDF}{CCDF}{complementary cumulative distribution function}
\newacronym{CDF}{CDF}{cumulative distribution function}
\newacronym{PMF}{PMF}{probability mass function}
\newacronym{IM}{IM}{index modulation}
\newacronym{JCAS}{JCAS}{joint communication and sensing}
\newacronym{ATSC}{ATSC}{advanced television systems committee}

\begin{document}
\title{PAPR-optimized OFDM Design for\\ Opportunistic Communications and Sensing}
\author{\IEEEauthorblockN{Getuar Rexhepi\orcidlink{0009-0002-3268-522X}~\IEEEmembership{Student Member,~IEEE}, Kengo Ando\orcidlink{0000-0003-0905-2109}~\IEEEmembership{Graduate Student Member,~IEEE,} and\\
Giuseppe~Thadeu~Freitas~de~Abreu\orcidlink{0000-0002-5018-8174}~\IEEEmembership{Senior Member,~IEEE} \\
emails: {\tt[grexhepi,kando,gabreu]@constructor.university} \\
}
\IEEEauthorblockN{School of Computer Science and Engineering, Constructor University Bremen, Germany}
}
\maketitle 

\begin{abstract}
We consider the problem of \ac{PAPR} reduction in \ac{OFDM} systems via optimized sparsification of \ac{TR} aimed at freeing resources for the opportunistic operation of co-existing communication and sensing systems.
In particular, we propose a novel \ac{TR}-optimization method in which the minimum number of effectively used \acp{PRT} required to satisfy a prescribed \ac{PAPR} level in a primary system is found, leaving the remaining \acp{PRT} free to be opportunistically utilized by a secondary system and other functionalities, such as \ac{JCAS}, \ac{IM} and \ac{CR}.
The proposed method relies on an $\ell_0$ norm regularization approach to penalize the number of \acp{PRT}, leading to a problem convexized via \ac{FP}, whose solution is shown to ensure that the prescribed \ac{PAPR} is achieved with high probability with a smaller number of \acp{PRT} than \ac{SotA} methods.
The contribution can be seen as a mechanism to enable the opportunistic coexistence of systems with adjacent functionalities in presence of existing \ac{OFDM}-based systems.
\end{abstract}

\glsresetall\section{Introduction}
\Ac{OFDM} is a modulation technique that is widely adopted in modern wireless communication standards, due to its robustness against multipath fading, its ability to mitigate inter-symbol interference, and the easiness of integration with multi-access schemes.
A known issue with \ac{OFDM}, however, is that it may suffer from high \ac{PAPR} as a consequence of the combination of a large number of subcarriers resulting from \ac{IFFT} signal processing.
High \ac{PAPR} is indeed one of the most undesirable drawbacks of \ac{OFDM} systems, as it diminishes \ac{SQNR} of both \ac{ADC} and \ac{DAC}, greatly affecting the efficiency of transmit power amplifiers.
The aim of the \ac{PAPR} reduction techniques is to decrease the signal's peak in the time domain, without corrupting its spectrum.
During the last two decades, several \ac{PAPR} reduction techniques have been proposed to mitigate high \ac{PAPR} in \ac{OFDM} systems, including coding schemes, clipping, constellation shaping,  \ac{SLM} and \ac{TR} \cite{Tellado1998, JiangTB2008, RahmatallahCST2013, SelahattinTVT2019, TarakTB2021}.
Among these techniques, the \ac{TR} method is a particularly simple alternative in which a small percentage of the total number of subcarriers, referred to as \acp{PRT}, are not used for data-transmission, but instead assigned random signals designed to reduce the \ac{PAPR} of the aggregate transmitted signal.
As such, the primary goal of the \ac{TR}-based \ac{PAPR} reduction method is to find the optimal complex symbols assigned to the \ac{PRT} subcarriers such that the \ac{PAPR} of the transmitted signal is minimized.
Originally formulated as a convex optimization problem, the \ac{TR} method can be easily casted as a \ac{LP} problem \cite{Tellado1998} and therefore solved very efficiently.

Thanks to its simplicity and effectiveness, the \ac{TR} method has been widely adopted by different standards, such as \ac{DVB-T2} \cite{DVB-T2Standard2012}, \ac{DVB-NGH}, and the {ATSC} 3.0 standard \cite{ATSC3_Standard2024} of the advanced television systems committee.
It is typical, in \ac{PAPR} reduction methods exploiting the \ac{TR} approach, to assume that all the \ac{PRT} subcarriers are required in order to achieve the desired target \ac{PAPR}.
From an optimization-theoretical perspective, on the other hand, it is not hard to infer that under many signal realizations that is not the case, since it is known that the \ac{PDF} of \ac{PAPR} levels is well concentrated, with a relatively thin upper tail \cite{OchiaiTC2001}.

In light of the above, and motivated by opportunistic techniques such as \ac{OFDM}-based \ac{CR} \cite{MahmoudWLM2009}, \ac{JCAS} \cite{KeskinTSP2023} and \ac{IM} \cite{LiTWC2020}, we propose in this paper a novel \ac{TR} method to minimize instantaneous the number of \ac{PRT} subcarriers effectively used for \ac{PAPR} reduction, such that the unused spectrum can be opportunistically used by secondary systems for other purposes, such as communication and sensing \cite{LabibRSB2017,ParmentarRadCon2019,LiuJSAC2022}, such that the primary system's performance is not affected.
The proposed method is based on the minimization of the $\ell_0$ norm of the signal vector designed to be transmitted over the \acp{PRT}, constrained by a maximum transmitted power and by the desired target \ac{PAPR}.
Simulation results indicate that the proposed method achieves the same \ac{PAPR} reduction performance as \ac{SotA} alternatives for a given prescribed \ac{PAPR}, while using a smaller number of \ac{PRT} subcarriers, thus enabling the opportunistic exploitation of those unused subcarriers by secondary systems.

\vspace{-1ex}
\section{System Model}

\subsection{OFDM Signal Model and PAPR}

Consider an \ac{OFDM} system with $N$ subcarriers, such that the time-domain transmit signal $\mathbf{x}\!\in\!\mathbb{C}^{N \times 1}$ corresponding to an information vector of complex symbols $\bm{x}\!\in\!\mathbb{C}^{N\times 1}$ can be expressed as
\begin{equation}
\label{eq:OFDM}
\mathbf{x} = \mathbf{F}_N^{\mathrm{H}}\bm{x},
\end{equation}
where $\mathbf{F}_N$ denotes the normalized \ac{DFT} matrix of size $N$.

The associated instantaneous \ac{PAPR} measures the maximum power of the signal relative to its average, and is defined as \cite{DinurTC2001}
\begin{equation}
\label{eq:PAPR}
\rho(\mathbf{x}) \triangleq \frac{\|\mathbf{x}\|_{\infty}^2}{\frac{1}{N}\| \mathbf{x} \|_2^2},
\end{equation}
where $\| \cdot \|_{\infty}$ and $\| \cdot \|_2$ denote the $\ell_{\infty}$ and $\ell_2$ norm operations, respectively.

\subsection{SotA Design of Reserved Tones for PAPR Reduction}

The \ac{TR} method relies on the reservation of a certain number of subcarriers, which instead of carrying information symbols are instead allocated dummy signals designed to minimize the \ac{PAPR} reduction of \ac{OFDM} transmit signals.
Mathematically, this corresponds to saying that the signal $\bm{x}$ in equation \eqref{eq:OFDM} can be decomposed as
\begin{subequations}
\label{eq:TR_x}
\begin{equation}
\label{eq:TR_x_FD}
\bm{x} = \bm{d} + \bm{r},
\end{equation}
such that, equivalently,
\begin{equation}
\label{eq:TR_x_TD}
\mathbf{x} = \mathbf{d} + \mathbf{r},
\end{equation}
where $\bm{d}\!\in\!\mathbb{C}^{N\times 1}$ and $\mathbf{d}\!\in\!\mathbb{C}^{N\times1}$ correspond to data communication symbols, while $\bm{r}\!\in\!\mathbb{C}^{N\times 1}$ and $\mathbf{r}\!\in\!\mathbb{C}^{N\times1}$ correspond to reserved tone symbols, and the equivalence between equations \eqref{eq:TR_x_FD} and \eqref{eq:TR_x_TD} follows from the linearity of the \ac{IFFT} and the fact thats $\mathbf{d} = \mathbf{F}_N^{\mathrm{H}}\bm{d}$ and $\mathbf{r} = \mathbf{F}_N^{\mathrm{H}}\bm{r}$.
\end{subequations}

In order to avoid the collision of data and reserved tone symbols, a mutually-exclusive subcarrier allocation is needed.
Let $\mathcal{N}\!\triangleq\!\{1,\dots,N\}$ denote the set of all subcarrier indices, such that the sets of data and reserved tone subcarrier indices can be respectively denoted by $\mathcal{D}\subseteq \mathcal{N}$ and $\mathcal{R}\subseteq \mathcal{N}$, with $\mathcal{N} = \mathcal{D}\cup\mathcal{R}$ and $\mathcal{D}\cap\mathcal{R} = \emptyset$.
Finally, let us also define the corresponding cardinalities $N_\mathrm{D} \triangleq |\mathcal{D}|$ and $N_\mathrm{R} \triangleq |\mathcal{R}|$, such that $N \!=\! N_\mathrm{D} + N_\mathrm{R}$.

From the above, it follows that the data and reserved tone signal vectors are such that their entries satisfy
\begin{equation}
d_n  =  0, \forall\, n \in \mathcal{R}\quad\text{and}\quad
r_n  = 0, \forall\, n \in \mathcal{D}.
\end{equation}

For the sake of convenience, we shall then introduce the notation
$\bm{r}\in\mathbb{C}^\mathcal{R}$ to indicate that elements of the sparse vector $\bm{r}$ corresponding to the indices in $\mathcal{R}$ take on complex numbers, while the remaining entries are zero, such that the optimization problem for \ac{PAPR} minimization can then be concisely formulated as
\begin{subequations}
\label{eq:PAPRMinimize}
\begin{align}
& \underset{\bm{r}\in\mathbb{C}^\mathcal{R}}{\text{minimize}}
& & \rho(\bm{r};\bm{d}) \\
& \text{subject to}
\label{eq:Original_PAPRConstraint}
& & \|\bm{r}\|_2^2 \leq P_\mathrm{\max},
\end{align}
\end{subequations}
where $P_\mathrm{\max}$ is a power constraint, and we slightly upgraded the notation to indicate that the \ac{PAPR} is, from the viewpoint of the optimization problem, a function of the sparse \ac{PRT} vector $\bm{r}$, parameterized by the data vector $\bm{d}$.

Since problem~\eqref{eq:PAPRMinimize} is not convex, it is difficult to solve efficiently, which motivates relaxations found in \ac{SotA} methods, such as the one proposed \cite{BulusuITBC2018} where the problem is reformulated into the following \ac{QCQP}

\quad\\[-6ex]
\begin{subequations}
\label{eq:Sota}
\begin{align}
& \underset{\bm{r}\in\mathbb{C}^\mathcal{R}}{\text{minimize}}
& & \| \mathbf{d} + \mathbf{F}_N^{\mathrm{H}} \bm{r} \|_{\infty}^2 \\
& \text{subject to}
\label{eq:SotA_PAPRConstraint}
& & \|\bm{r}\|^2_{\infty} \leq \frac{\Omega}{N-N_\mathrm{R}} \|\bm{d}\|_2^2,
\end{align}
\end{subequations}
where $\Omega$ is a given power level gap constant \cite{BulusuITBC2018}.

Inspecting equation \eqref{eq:Sota} in light of equations \eqref{eq:PAPR} and \eqref{eq:TR_x}, it can be seen that the objective function of the \ac{QCQP} optimization problem corresponds only to the numerator of the \ac{PAPR}, which implies that the \ac{SotA} approach does not attempt to reduce \ac{PAPR} itself, but merely the peak power.
In addition, it is visible from constraint \eqref{eq:SotA_PAPRConstraint} that the method does not enforce a hard limit on the power of the \ac{PRT} vector $\bm{r}$, as originally in constraint \eqref{eq:Original_PAPRConstraint}, but rather attempts only to limit it to a fraction of that of the data symbol vector $\bm{d}$.

It is evident from all the above, including the formulation in equation \eqref{eq:Sota}, that there is no mechanism optimize the number of \ac{PRT} subcarriers used in \ac{PAPR} reduction.
In other words, the \ac{SotA} scheme both assumes that the cardinality $N_\mathrm{R}$ of the set $\mathcal{R}$ of \ac{PRT} subcarriers is given, and that all such subcarriers are needed to achieve the target \ac{PAPR}.

\begin{figure*}[t]
  \centering
  \includegraphics[width=\textwidth]{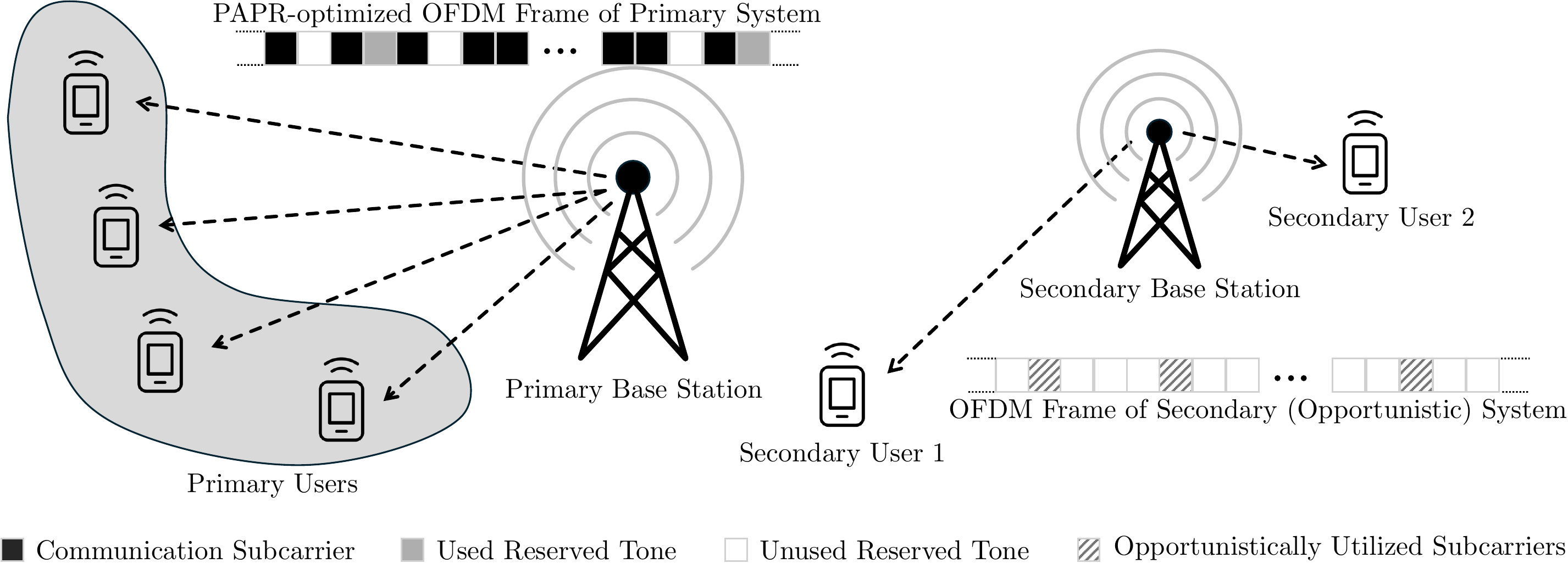}
  \caption{OFDM structure with reserved tones for PAPR reduction.}
  \label{fig:OFDM}
\end{figure*}

\section{Proposed PAPR Reduction Scheme}

In view of the above, we propose in the sequel an alternative formulation of \ac{RT}-based \ac{PAPR} problem, which not only more frontally addresses the \ac{PAPR} reduction objective, but also enables a better control of the actual utilization of the \ac{PRT} subcarriers, in the sense that it leads to the smallest number of tones to achieve the prescribed \ac{PAPR}.

Referring to the illustration shown in Figure~\ref{fig:OFDM}, that key idea is to maximize the effective use of spectrum by utilizing \ac{PRT} subcarriers only to the extent that they are needed in order to achieve the desired \ac{PAPR} level, leaving therefore tones that are eventually not needed to be utilized opportunistically by secondary systems for other functionalities.

\subsection{Problem Formulation}

Straightforwardly, consider the problem
\begin{subequations}
\label{eq:Prop}
\begin{align}
& \underset{\bm{r}\in\mathbb{C}^\mathcal{R}}{\text{minimize}}
&& \|\bm{r}\|_0 \\
& \text{subject to}
&& \|\bm{r}\|_{2}^2 \leq \text{P}_{\max} \\
\label{eq:Prop_PAPRConstraint}
&&& \rho(\bm{r};\bm{d}) \leq \rho^*,
\end{align}
\end{subequations}
where $\|\cdot\|_0$ denotes the $\ell_0$-norm operator and $\rho^*\!\in\!\mathbb{R}$ denotes a target \ac{PAPR}.

Problem \eqref{eq:Prop} is not convex due both to the $\ell_0$-norm in the objective function and the \ac{PAPR} constraint \eqref{eq:Prop_PAPRConstraint}.
In order to convexize the objective, consider the following continuous and differentiable approximation of the $\ell_0$-norm \cite{IimoriOJCS2021}
\begin{equation}
\label{eq:Approx}
\|\bm{r}\|_0 \approx \sum_{n=1}^N \frac{|r_n|^2}{|r_n|^2+\alpha} = N - \sum_{n=1}^{N} \frac{\alpha}{|r_n|^2+\alpha},
\end{equation}
where $\alpha\!\in\!\mathbb{R}$ denotes a parameter controlling the tightness of the approximation. 

Then, let us convexize the affine-over-convex ratios in equation \eqref{eq:Approx} via the \ac{FP} approach \cite{ShenTSP2018}, from which we obtain
\begin{equation}
\label{eq:FP}
\|\bm{r}\|_0 \approx N - \left( \sum_{n=1}^{N} 2 \gamma_n \sqrt{\alpha} - \gamma_n^2 |r_n|^2 + \gamma_n^2 \alpha \right),
\end{equation} 
where $\gamma_n$ is the \ac{FP} auxiliary variable given by
\begin{equation}{\label{eq:gamma}}
\gamma_n = \frac{\sqrt{\alpha}}{|r_n|^2 + \alpha}.
\end{equation}

We clarify that under the \ac{FP} framework, the quantities $\alpha$ and $\gamma_n$ are kept \underline{constant} during each complete run of the procedure, being updated subsequently.
As a result, within each run of the optimization solver, only the third term in equation \eqref{eq:FP} depends on the variable $\bm{r}$, such that for the purpose of the optimization problem we have
\vspace{-0.5ex}
\begin{equation}
\label{eq:ApproxFinal}
\|\bm{r}\|_0 \equiv  \sum_{n=1}^{N} \gamma_n^2 |r_n|^2 = \bm{r}^\text{H}\mathbf{\Gamma}\bm{r},
\vspace{-0.5ex}
\end{equation}
where it is implicit that $\mathbf{\Gamma}\triangleq \mathrm{diag}([\gamma_1^2,\cdots,\gamma_N^2])$.

Constraint~\eqref{eq:Prop_PAPRConstraint} can also be convexized similarly to the above.
To that end, let us first rewrite the inequality as
\vspace{-0.5ex}
\begin{equation}
\frac{\frac{1}{N}\|\mathbf{x}\|_2^2}{\|\mathbf{x}\|_{\infty}^2} \geq \frac{1}{\rho^*},
\vspace{-0.5ex}
\end{equation}
such that we again obtain via \ac{FP} \cite{ShenTSP2018}
\vspace{-0.5ex}
\begin{equation}
2 \Re\{\boldsymbol{\zeta}\mathbf{x}\}  - \boldsymbol{\zeta}^{\mathrm{H}}\boldsymbol{\zeta} \|\mathbf{x}\|_{\infty}^2  \geq \frac{N}{\rho^*},
\vspace{-0.5ex}
\end{equation}
where $\boldsymbol{\zeta}\!\in\!\mathbb{C}^{N\times 1}$ is an \ac{FP} auxiliary vector, given by
\begin{equation}\label{eq:zeta}
\boldsymbol{\zeta} = \frac{\mathbf{x}}{\|\mathbf{x}\|_{\infty}^2}.
\end{equation}

From all the above, the optimization problem in equation \eqref{eq:Prop} can be reformulated into the convex problem
\begin{subequations}
\label{eq:Prop2}
\begin{align}
& \underset{\bm{r}\in\mathbb{C}^\mathcal{R}}{\text{minimize}}
&& \bm{r}^\text{H}\mathbf{\Gamma}\bm{r} \\
& \text{subject to}
&& \|\bm{r}\|_{2}^2 \leq {P}_\mathrm{\max} \\
&&& 2 \Re\{\boldsymbol{\zeta}\mathbf{x}\}  - \boldsymbol{\zeta}^{\mathrm{H}}\boldsymbol{\zeta} \|\mathbf{x}\|_{\infty}^2  \geq \frac{N}{\rho^*},
\end{align}
which can be solved efficiently.
\end{subequations}

\subsection{Algorithmic Considerations}

Before we move on to assess the performance of the \ac{PAPR} scheme introduced above, let us address some relevant algorithmic considerations about the proposed method.
For starters, we emphasize that the \ac{FP} approach employed requires an initial solution to be kick-started, as evidenced by the auxiliary quantities $\mathbf{\Gamma}$ and $\boldsymbol{\zeta}$, which are computed and kept constant for a given iteration, based on the result of a previous iteration.
In what follows, it will be assumed that the \ac{SotA} approach of \cite{BulusuITBC2018} will be used to initialize the proposed algorithm.

Secondly, let us also call attention to the fact that unlike the formulation in equation \eqref{eq:Sota}, where total transmit power is left unconstrained and where no target \ac{PAPR} is specified, problem \eqref{eq:Prop2} has two absolute constraints related to these system parameters, such that a solution of the problem may occasionally not be feasible.
And since in such cases the \ac{FP}-based proposed algorithm naturally yields the initial solution as its own output, it can ultimately be said that the proposed technique is fundamentally a refinement stage complementary to the \ac{SotA} method of \cite{BulusuITBC2018}.

Finally, we remark that due to the approximation of the $\ell_0$-norm in equation \eqref{eq:Approx}, solutions of problem \eqref{eq:Prop2} may contain very small but non-zero elements, which can be zero-forced by comparison with a given threshold $\varepsilon\!\in\!\mathbb{R}$, namely

\begin{algorithm}[H]
\caption{: \ac{TR}-based \ac{PAPR} Reduction Scheme via \ac{FP}}
\label{Alg:prop}
\begin{algorithmic}[1]
\vspace{-0.4ex}
\Statex \hspace{-4.5ex} {\bf Input:} Time-domain data vector $\mathbf{d}\!\in\!\mathbb{C}^{N\times 1}$; sets of data and
\Statex \hspace{-4ex}\ac{PRT}  subcarrier indices $\mathcal{D}$ and $\mathcal{R}$, respectively; maximum TX
\Statex \hspace{-4.5ex} power $P_\mathrm{\max}$; \ac{PAPR} threshold $\rho^*$; accuracy parameter $\alpha$ for
\Statex \hspace{-4.5ex} $\ell_0$-norm approximation; and \ac{PRT} tolerance $\varepsilon$.
\Statex \hspace{-4.5ex} {\bf Output:} Optimized sparse reserved tone vector $\bm{r}\!\in\!\mathbb{C}^\mathcal{R}$
\vspace{.2ex}\hrule \vspace{.6ex}
\vspace{.2ex}\hrule \vspace{.6ex}
\Statex \hspace{-4.5ex} {\bf Initialization:} Solve problem \eqref{eq:Sota} to obtain initial vector $\bm{r}$
\vspace{.2ex}\hrule \vspace{.6ex}
\State {\bf repeat} until convergence of $\bm{r}$
\State Construct $\mathbf{\Gamma}\triangleq \mathrm{diag}([\gamma_1^2,\cdots,\gamma_N^2])$ with $\gamma_n$  as in equation equations \eqref{eq:gamma}
\State Construct $\mathbf{x}$ using equations \eqref{eq:OFDM} and \eqref{eq:TR_x}
\State Construct $\boldsymbol{\zeta}$ via equation \eqref{eq:zeta}
\State Update $\bm{r}$ by solving problem \eqref{eq:Prop2}
\State Enforce sparsity via equation \eqref{eq:threshold}
\end{algorithmic}
\end{algorithm}

\vspace{-4ex}
\begin{equation}
\label{eq:threshold}
r_n =
\begin{cases}
 r_n, &  |r_n| \geq \varepsilon, \,\, n\in \mathcal{R}\\
 0, &  \text{otherwise}. \\
\end{cases}
\end{equation}

A summary of the proposed scheme in the form of a pseudo-code is given above in Algorithm \ref{Alg:prop}.

\vspace{-1ex}
\section{Simulation Results}

\begin{figure}[H]
\centering
\includegraphics[width=0.5\textwidth]{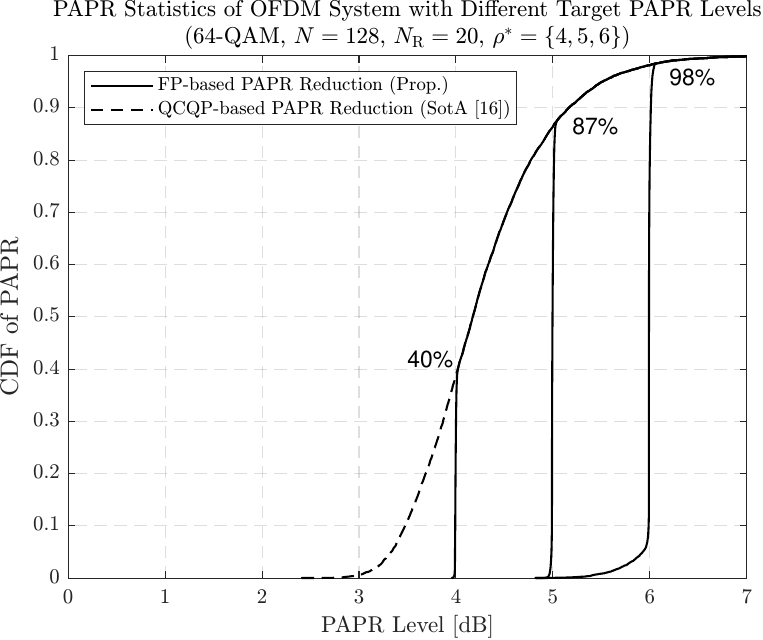}
\caption{CDFs of PAPRs achieved by proposed and SotA PAPR reduction methods, with different target PAPRs.}
\label{fig:CDF}
\end{figure}

In this section, we evaluate the performance of proposed method via computer simulations. 
We start with a plot of the \acp{CDF} of the achieved instantaneous \ac{PAPR} for a system with $N = 128$ subcarriers, out of which $R = 20$ are reserved for \ac{PAPR} reduction.
The results, shown in Figure \ref{fig:CDF}, compare the \acp{CDF} achieved with the proposed algorithm using $\alpha = 10^{-4}$, $\varepsilon = 7 \cdot 10^{-4}$ and $\Omega = 10$, for 3 distinct target \ac{PAPR} levels, namely, $\rho^* = 4, 5$ and $6$ dB.
The figure demonstrates that the proposed scheme is very effective in reducing the number of \ac{PRT} subcarriers from $20$ to only $12$, $6$ and $3$ tones, respectively, achieving the target \ac{PAPR} $40\%$, $87\%$, and $98\%$ of the time, respectively, as indicated by the sharp ascent of the \acp{CDF}.
\vspace{-1ex}

\begin{figure}[H]
\centering
\includegraphics[width=0.5\textwidth]{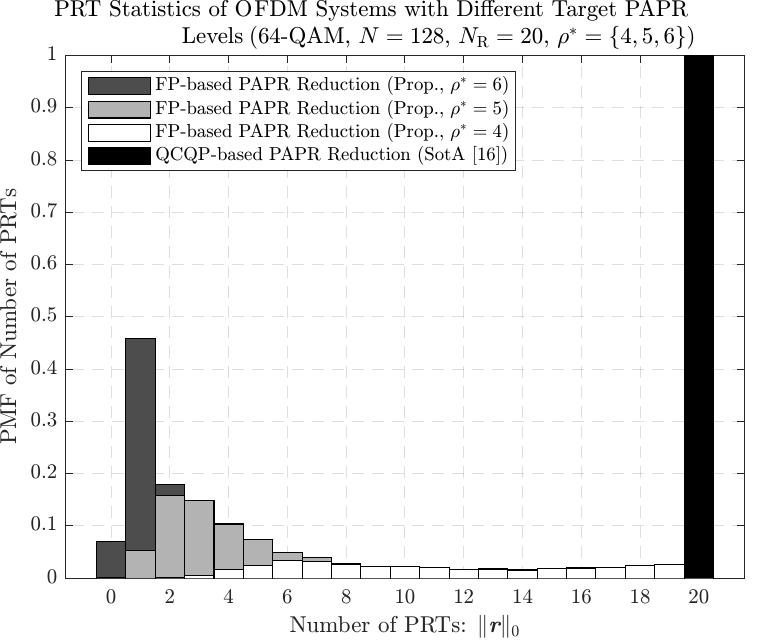}
\caption{Distributions of number of \acp{PRT} required by proposed and SotA PAPR methods in order to achieve target PAPR levels.}
\label{fig:PAPR}
\vspace{1ex}
\includegraphics[width=0.5\textwidth]{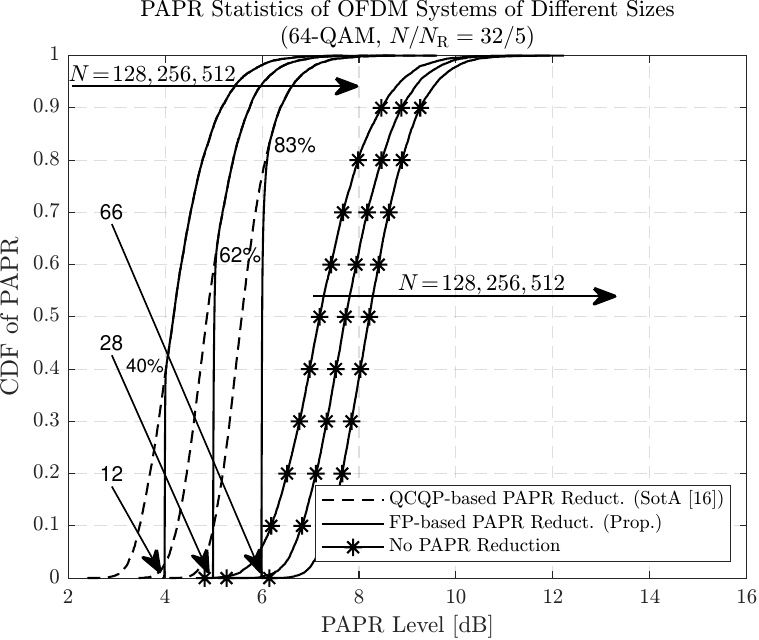}
\caption{CDFs of PAPRs achieved by OFDM systems of different sizes without and with proposed and SotA PAPR reduction methods, with different target PAPRs.}
\label{fig:CDFMulti}
\vspace{-2ex}
\end{figure}

Next, we compare in Figure \ref{fig:PAPR} the \acp{PMF} of the number of \acp{RT} required to achieve the latter target \acp{PAPR}, namely, $\rho^* = 4, 5$ and $6$ dB.
It can be seen, once again, that indeed the proposed method is capable of freeing, with high probability, a large number of reserved tones, without sacrificing the desired \ac{PAPR} level.
In particular, it is observed that as $\rho^*$ grows, the further to the left and the more concentrated is the center of mass of the \ac{PMF}.

Finally, we compare in Figure \ref{fig:CDFMulti} the \acp{CDF} of the \ac{PAPR} levels achieved by \ac{OFDM} systems of different sizes, without and with the proposed and SotA PAPR-reduction methods, with different target PAPRs and employing the same ratio between \acp{PRT} and total number of carriers.
In this figure, it can be seen that the target PAPR levels are much lower than those typically observed in the corresponding \ac{OFDM} system without \ac{RT}-based \ac{PAPR} reduction.

The results indicates that as the system size grows, the easier it is for the proposed scheme to ensure, with a high probability, a PAPR level much lower than that of a non-optimized system, with the same fraction of \acp{PRT} relative to the total number of subcarriers.
This implies, therefore, that the proposed scheme is such that in \ac{OFDM} systems with a large number of subcarriers, a very significant \ac{PAPR} reduction can be achieved with only a very small number of reserved tones, with high probability.

Unfortunately the \ac{SotA} scheme of \cite{BulusuITBC2018} utilized here to initialize the proposed method is too computationally demanding to allow simulation results with the typical parameterization of \ac{OFDM} which has thousands of subcarriers.
In order to more directly demonstrate the aforementioned feature of the proposed scheme, we will, in a follow up work (or, time allowing, in the camera-ready version of this article), introduce a lower-complexity initialization alternative that allows for large scale simulations.

\section{Conclusions}
We proposed a novel \ac{PAPR} reduction in scheme for \ac{OFDM} systems, suitable to enable opportunistic utilization of radio spectrum.
In the proposed method, the minimum number of effectively used \acp{PRT} required to satisfy a prescribed \ac{PAPR} level in a primary system is found, leaving the remaining \acp{PRT} free to be opportunistically utilized by a secondary system and other functionalities, such as \ac{JCAS}, \ac{IM} and \ac{CR}.
The proposed method relies on an $\ell_0$ norm regularizer that penalizes the number of \acp{PRT}, leading to a problem convexized via \ac{FP}, whose solution was shown to ensure that the prescribed \ac{PAPR} is achieved with high probability with a smaller number of \acp{PRT} than \ac{SotA} methods.
The contribution can be seen as a mechanism to enable the opportunistic coexistence of systems with adjacent functionalities in presence of existing \ac{OFDM}-based systems.

\bibliographystyle{IEEEtran}
\bibliography{reference}

\end{document}